 \documentstyle[epsf,twocolumn,prl,aps]{revtex}

\begin{document}
\draft
%\preprint{HEP/123-qed}
% comment out the following line for single column output
 \twocolumn[\hsize\textwidth\columnwidth\hsize\csname @twocolumnfalse\endcsname

\title{Mott-Peierls Transition in the extended Peierls-Hubbard model} 
\author{Eric Jeckelmann}
\address{Department of Physics and Astronomy, 
University of California, Irvine, CA 92697.}
\date{\today}
\maketitle
\begin{abstract}
The one-dimensional extended Peierls-Hubbard model is studied at several
band fillings using the density matrix renormalization group method.
Results show that the ground state evolves from a Mott-Peierls 
insulator with a correlation gap at half-filling
to a soliton lattice with a small band gap away from half-filling. 
It is also confirmed that the ground state of the Peierls-Hubbard model
undergoes a transition to a metallic state at finite doping.
These results show that electronic correlations effects should be taken
into account in theoretical studies of doped polyacetylene.
They also show that a Mott-Peierls theory could explain 
the insulator-metal transition observed in this material. 
\end{abstract}
\pacs{71.10.Fd, 71.30.+h, 71.10.Pm, 71.20.Rv}

% comment out the following line for single column output
 ]

\narrowtext

Since the discovery of the metallic phase of doped polyacetylene,
this material has been extensively studied~\cite{hkss88,kiess},  
but the mechanism of the insulator-metal transition 
observed upon doping is still poorly understood.
It is known that both the Peierls instability and
electronic correlations play a fundamental role in the formation and
the properties of the insulating phase~\cite{bcm92}
and thus undoped polyacetylene is a Mott-Peierls insulator~\cite{ovc87}. 
Therefore, ten years ago, Baeriswyl, Carmelo and Maki~\cite{bcm87} 
proposed that the insulator-metal transition 
was also driven by the interplay of electron-electron
and electron-phonon interactions. 
Within the restricted Hartree-Fock approximation they have
shown the possibility of such a Mott-Peierls insulator-metal
transition in the Peierls-Hubbard model, 
which is the simplest model of polyacetylene including both 
interactions.
Recently, several works using sophisticated numerical many-body methods, 
such as the Gutzwiller variational wavefunction~\cite{jb94} and
quantum Monte Carlo (QMC) simulations~\cite{tak96},
have confirmed the occurrence of an insulator-metal
transition in this model. 
On the other hand, Wen and Su, which have applied the
the density matrix renormalization group (DMRG) technique to this problem,
have disputed the existence of this transition~\cite{wen96}. 
Hartree-Fock~\cite{bj92} and QMC~\cite{tak96} simulations 
have also shown that the nearest-neighbor Coulomb repulsion 
opposes and can prevent the formation of a metallic state 
in the extended Peierls-Hubbard model.

As an attempt to clarify this issue
I have studied the properties of the extended Peierls-Hubbard
model with parameters leading to a Mott-Peierls
insulating ground state at half-filling. 
Accurate ground states and gaps are obtained for open chains 
up to 200 sites and different band fillings using
the DMRG method~\cite{whi92} and finite size effects have been
carefully analyzed. 
Results show that the ground state of a Mott-Peierls insulator 
evolves to a soliton lattice upon doping.
This soliton lattice is qualitatively similar to the 
ground state predicted by simple electron-phonon models~\cite{ssh80,sol},
but both the gap and the amplitude of the lattice distortion
decrease faster in the extended Peierls-Hubbard model than in these models 
when the doping increases.
An insulator-metal transition occurs at a finite doping concentration
in the absence of nearest-neighbor electron-electron interaction, 
in agreement with previous studies of the Peierls-Hubbard 
model~\cite{bcm87,jb94,tak96}.
These results demonstrate that electronic correlations effects
are important and should be taken into account in theoretical studies 
of doped polyacetylene.
They also confirm that a Mott-Peierls theory~\cite{ovc87,bcm87,jb94} 
could explain
the insulator-metal transition observed in polyacetylene.

The one-dimensional extended Peierls-Hubbard model is defined by the 
Hamiltonian 
%\begin{equation}
\begin{eqnarray}
H =  && {1 \over {4 \pi t \lambda}} \sum_{\ell} \Delta^{2}_{\ell} - 
{2 P \over \pi} \ \sum_{\ell}  (-1)^\ell \Delta_{\ell} \  \nonumber \\
&& - \sum_{\ell\sigma} \left(t-(-1)^\ell {\Delta_{\ell} \over 2}\right) 
 \left (c^{+}_{\ell+1\sigma}
c_{\ell\sigma} + c^{+}_{\ell\sigma} c_{\ell+1\sigma} \right )
\nonumber \\
&& + U \sum_{\ell} n_{\ell\uparrow} n_{\ell\downarrow}  
+ V \sum_{\ell}  n_{\ell} n_{\ell+1} \ .
\label{ham}
%\end{equation}
\end{eqnarray}
The operators $c^{+}_{\ell\sigma} (c_{\ell\sigma})$ create (destroy) an 
electron of spin $\sigma$ at site $\ell$, $n_{\ell \sigma} = 
c^{+}_{\ell\sigma} c_{\ell\sigma}$ and {$n_{\ell} = n_{\ell \uparrow} 
+ n_{\ell \downarrow}$}. 
$t$ is the resonance integral for an undistorted lattice and fixes the
energy scale, $\lambda$ is the electron-phonon coupling constant,
$U$ and $V$ are the on-site and nearest-neighbor Coulomb repulsion.
As this model  has an electron-hole symmetry, only hole doping is examined. 
The doping rate $y$ is defined as the fraction of electrons removed from a
neutral chain (which corresponds to a half-filled band).
The usual dimerization order parameter $\Delta_{\ell}$ describes
the lattice degrees of freedom.
A linear term with constant $P$ is explicitly included in the lattice 
elastic energy instead of the constraint on the dimerization order 
parameter used in previous works~\cite{jb94,wen96}
in order to reduce the average bond length variation.
The value of $P > 0$ is determined by the condition
that the linear term in the elastic energy equals zero
in the ground state configuration at half-filling.
The lattice dynamics is completely neglected in this approach and
the electron-phonon interaction is taken into account only through
the coupling between electrons and a classical lattice relaxation. 

To determine the ground state of the Hamiltonian (\ref{ham}), 
one has to find both the lattice configuration $\{\Delta_{\ell}\}$ 
and the electronic wavefunction which minimize the total energy.
Using a finite system DMRG algorithm~\cite{whi92}, 
one can compute the electronic 
ground state, its energy and the gradient of this energy 
(thanks to the Hellmann-Feynman theorem)
for any given lattice configuration, 
and thus perform the minimization of the total energy with respect to
lattice degrees of freedom $\{\Delta_{\ell}\}$~\cite{jb94,wen96}.
In principle, direct electronic excitations can also be obtained by 
calculating excited states of the electronic part of the Hamiltonian 
(\ref{ham}) for a fixed lattice configuration.
Unfortunately, while the DMRG method gives excellent results 
for ground states,  
it is more difficult to obtain results for specific excited states. 
Therefore, I have only calculated charge gaps, which
can easily be obtained from ground state energies for different band 
fillings ~\cite{jb94}.
The charge gap is believed to be equal to the lowest optical
absorption energy in the thermodynamic limit of the Peierls-Hubbard 
model ($V=0$). 
In the extended Peierls-Hubbard model ($V \neq 0$) the relation 
between charge
gap and optical gap is not known precisely but in this work I have assumed
that both quantities are roughly equivalent in the thermodynamic limit.
  
All calculations have been carried out for several chain lengths 
up to 200 sites and results have always been extrapolated to 
an infinite chain. 
Only open chains are considered because the DMRG method performs much
better in this case than for periodic boundary conditions.
Computations have been performed so that 
numerical errors on the ground state dimerization parameter
$\Delta_{\ell}$ are smaller than $10^{-3} t$.   
Numerical errors on gap values are estimated to be less than $10^{-2} t$ 
at half filling and around $10^{-3}t$ away from half filling.
All these estimations of the accuracy are based on an analysis
of the behavior of DMRG results as a function of the number $m$ of 
quantum states kept per block. 
The largest  value of $m$ used in this work ranges from 80 for short 
chains (50 sites) at half filling to 400 for long chains (200 sites) 
away from half filling.
Truncation errors are typically between $10^{-6}$ and $10^{-7}$.
I have also checked the accuracy of DMRG calculations against exact
numerical results for long (up to 100 sites) non-interacting 
($U = V = 0$) chains and against exact results for the one-dimensional 
Hubbard model ($\Delta_\ell = 0$ and $V = 0$)~\cite {lie68}. 
An excellent agreement has been found in both cases. 

My results at half filling are in good agreement with results 
obtained previously with DMRG~\cite{dmrg}
and other many-body techniques like exact diagonalizations,
quantum Monte Carlo simulations, and variational methods~\cite{bcm92}.
They confirm that undoped polyacetylene is a Mott-Peierls insulator, 
which can be described with a reasonable accuracy by 
the extended Peierls-Hubbard models.
I have determined appropriate parameters for polyacetylene by comparing 
model predictions to experimental values for the optical gap 
at half-filling, the  

\vbox{
\begin{figure}[ht]
\epsfxsize=3.5 in\centerline{\epsffile{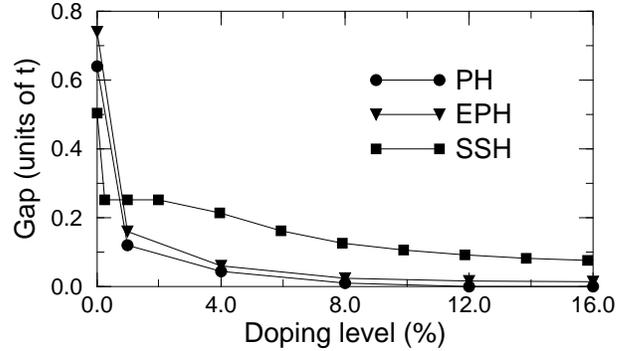}}
\caption{Optical gap (in units of $t$)
of the Peierls-Hubbard (PH) model, extended
Peierls-Hubbard (EPH) model and SSH model
as a function of doping.} 
\label{evo}
\end{figure}
}

\noindent
optical transition energies induced by photo-generated
neutral and charged solitons
and the neutral soliton spin density obtained from
magnetic resonance experiments~\cite{hkss88,kiess}.
This comparison shows that 
$\lambda = 0.1$, $U = 4V = 2.5 t$ and $t = 2.7 eV$ 
seem to be appropriate for polyacetylene 
in agreement with previous studies~\cite{bcm92,jb94}.

It is important to realize that in the Mott-Peierls regime
the optical gap at half-filling is essentially a correlation gap as in 
the one-dimensional Hubbard model~\cite{lie68}, although
the electron-phonon coupling and the Peierls instability
are responsible for features like the dimerization and
the existence of solitons.
For instance, DMRG calculations predict an optical gap $E_g = 0.74 t$
for the parameters mentioned above. 
As the gap is only $\sim 0.04t$ for $U=V=0$ and $\lambda = 0.1$, 
electronic correlations account for at least $94 \%$ of the gap.
Consequently, one expects this gap to be strongly reduced as soon as 
the system is doped because electronic correlations do not contribute to 
the formation of a gap away from half filling in the Hubbard model.
A strong experimental evidence for this reduction is the difference
between the gap at half filling ($1.8 eV$) and the energy of 
the optical transition induced by photo-generated charged solitons 
($0.45 eV$) which corresponds to the gap of a lightly doped chain in our 
simplified model.  

I have investigated the extended Peierls-Hubbard model at several
dopant concentrations up to $y = 16 \%$ 
in the Mott-Peierls regime.
In this regime the system evolves upon doping from the Mott-Peierls 
insulating state to a soliton lattice with a small gap.
The evolution of the optical gap upon doping is shown in Fig.~\ref{evo}
for the polyacetylene parameters mentioned previously.
As expected, the gap is strongly reduced to $\sim 0.17 t$
as soon as the system is doped.
The amplitude of the lattice distortion and the gap decrease 
with increasing doping but no transition to a metallic state is found 
up to the highest doping studied in this work ($y=16\%$). 
It is possible that a transition occurs at a higher doping
but this would not be relevant for the transition 
observed in polyacetylene 

\vbox{
\begin{figure}[ht]
\epsfxsize=3.5 in\centerline{\epsffile{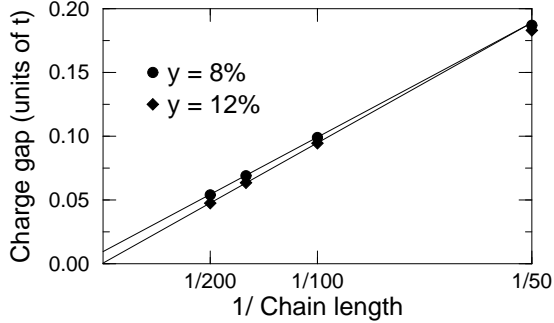}}
\caption{Charge gap (in units of $t$)
of the Peierls-Hubbard model
as a function of the inverse system size
for two different doping levels. 
Lines are linear extrapolations.}
\label{gapfig}
\end{figure}
}

\noindent
around $y = 6\%$.
Quantum Monte Carlo simulations~\cite{tak96} have also shown 
that the lattice
distortion survives at high doping for $V = U/2$.
For $y > 4 \%$, the amplitude of
the lattice distortion  $\Delta_\ell$ corresponds exactly 
to the value 
of the gap if both quantities are extrapolated 
to an infinite chain.
Therefore, away from half filling the gap is a band gap generated by the 
lattice modulation,
though electronic correlations contribute indirectly to its formation
because they increase the amplitude of the lattice distortion~\cite{jb94}.

The soliton lattice found in the doped extended Peierls-Hubbard model 
is qualitatively similar 
to the soliton lattice predicted by simple electron-phonon
models~\cite{sol}. 
However, it is important to realize that the evolution of the gap
and lattice distortion amplitude upon doping is 
different from the predictions of the Su-Schrieffer-Heeger~\cite{ssh80}
(SSH) model (shown in Fig.~\ref{evo} for $\lambda=0.2$).
To reproduce the results obtained in the extended Peierls-Hubbard model 
with the SSH model one would need to use an effective electron-phonon
coupling which decreases with increasing doping.
Moreover, such a doping-dependent parameter $\lambda(y)$
should change abruptly at
half-filling to reproduce the sudden disappearance of the correlation gap.
Finally, one notes that the energy 
scales involved in both approaches differ by an order of magnitude. 
For instance, my calculations indicate a gap of about 0.07 
$eV$ at $y=8\%$ while the SSH model predicts $0.4 eV$ \cite{jb94}. 
These results demonstrate that electronic correlations effects
in such one-dimensional systems
are not reproduced by simple single-electron models with 
an effective electron-phonon coupling,
contrary to a basic assumption of the SSH theory of conducting 
polymers~\cite{hkss88,ssh80}. 
Thus, the electron-electron interaction and electronic correlations effects
should be taken into account in theoretical studies
of doped polyacetylene.

In Fig.~\ref{evo} one can also see that the gap 
of the Peierls-Hubbard model with 
$\lambda = 0.1$ and $U = 2.5 t$ 
vanishes at a critical doping between 
$8$ and $12 \%$. 
These results confirm the existence of a transition to a metallic state 
at a finite 

\vbox{
\begin{figure}[ht]
\epsfxsize=3.5 in\centerline{\epsffile{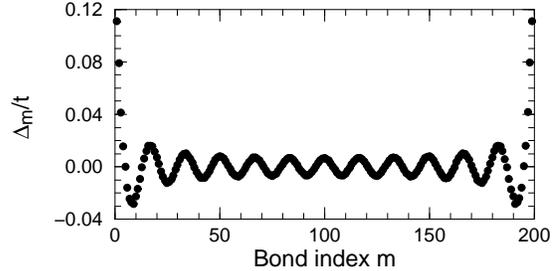}}
\caption{Lattice dimerization parameter $\Delta_m$ 
of the Peierls-Hubbard model
at $y=12\%$.}
\label{latfig}
\end{figure}
}

\noindent
doping~\cite{bcm87,jb94,tak96}.
They also
show that the electron-electron interaction can 
either support or oppose the Peierls instability 
away of half-filling depending on the parameters used, as predicted by 
the restricted Hartree-Fock approximation~\cite{bcm87,bj92}.
It should be noted that 
my numerical results agree quantitatively with 
those presented in Ref.~\cite{wen96} but additional calculations and an 
analysis of finite size and chain edge effects lead to a 
different conclusion. 
Fig.~\ref{gapfig} shows the charge gap as a function of the system size 
for $y = 8\%$ and $12 \%$. 
The value of the gap extrapolated to an infinite chain is clearly 
finite for $y = 8\%$ but vanishes for $y = 12 \%$ within numerical 
errors ($ \sim 10^{-3}t$).
The lattice dimerization parameter $\Delta_\ell$ is shown in
Fig.~\ref{latfig} for a 200-site chain at $y = 12 \%$.
The shape of $\Delta_\ell$ looks similar for $y = 8 \%$ except that
the amplitude is smaller at higher doping.
However, in the insulating phase ($y \leq 8\%$), 
the amplitude of the distortion in the middle of the chain
tends to a finite value as the chain length increases. 
This confirms that this lattice modulation 
is a genuine Peierls distortion.
On the other hand, 
in the metallic regime ($y \geq 12 \%$) the amplitude decreases 
as a power-law with an exponent $-0.66$ as the system size increases. 
This behavior demonstrates that the lattice distortion
is a chain edge effect with a very slow asymptotic decay.
The amplitude of these Friedel oscillations is known to decrease 
asymptotically as a power-law with an exponent -1 in a one-dimensional 
Fermi liquid, but density fluctuations are strongly affected by 
electron-electron interaction and the exponent can be as small as 
${-{1 \over 2}}$ in a Luttinger liquid \cite{egg95}.

In conclusion I have investigated the ground state
of the extended Peierls-Hubbard model in the Mott-Peierls
regime, which is appropriate for polyacetylene.
Results show that the ground state evolves from a Mott-Peierls 
insulator with a correlation gap at half-filling
to a soliton lattice with a small band gap away from half-filling. 
It is also confirmed that a insulator-metal transition
occurs in the Peierls-Hubbard model at a doping concentration
between $8$ and $12 \%$. 
These results clearly show that 
electronic correlations effects are important and 
should be taken into account in theoretical studies
of doped polyacetylene.
They also suggest that the primary mechanism of the 
insulator-metal transition in polyacetylene is the interplay 
between electron-phonon and electron-electron interactions, 
which induces a transition upon doping from an insulating 
state with a gap of $\sim 1.8 eV$ to a state with 
a gap which is two orders of magnitude smaller.
Obviously, this theoretical investigation of the properties
of an ideal, infinite and isolated chain cannot describe the 
properties of actual physical systems.
Understanding the insulator-metal transition of polyacetylene
will require the study of more realistic model including
lattice dynamics, interchain couplings, 
interaction with dopant ions and disorder~\cite{translist}.
In the future I plan to study such models with DMRG, particularly
the effects of quantum lattice dynamics.

%\acknowledgments
I thank S. White and D. Baeriswyl for helpful comments and discussion. 
I wish to acknowledge support from the  Swiss National Science Foundation. 
This work was also supported in part by the Campus Laboratory Collaborations 
Program of the University of California and by the NSF under Grant 
No. DMR-9509945. Some of the calculations were performed at
the San Diego Supercomputer Center.

%\appendix
%\section{Appendixes}

\end{document}